%Paper: hep-ph/9408211
%From: JPHALYO@wiswic.weizmann.ac.il
%Date: Tue, 2 Aug 1994 13:36:29 GMT

% Printing instructions:
%       This paper needs the macro package phyzzx.tex
%
\input phyzzx
\tolerance=1000
\sequentialequations
\def\rl{\rightline}

\def\r#1{$\bf#1$}

\def\t1{{\tilde 1}}

\def\F{\widetilde F}

\def\AEF{A.E. Faraggi}
\def\DVN{D. V. Nanopoulos}

\def\NPB#1#2#3{Nucl. Phys. B {\bf#1} (19#2) #3}
\def\PLB#1#2#3{Phys. Lett. B {\bf#1} (19#2) #3}

\def\PRT#1#2#3{Phys. Rep. {\bf#1} (19#2) #3}

\def\IJMP#1#2#3{Int. J. Mod. Phys. A {\bf#1} (19#2) #3}

\def\l{\langle}
\def\r{\rangle}

%\REF\WIT{E. Witten, \NPB{202}{82}{253}.}
\REF\SUGRA{H.P. Nilles, \PRT{110}{84}{1}; \DVN~ and Lahanas, \PRT{145}{87}{1}.}
\REF\GCM{
%H. P. Nilles, \PLB{115}{82}{193}; \NPB{217}{83}{366};
J. P. Deredinger, L. E. Ibanez and H. P. Nilles, \PLB{155}{85}{65};
M. Dine, R. Rohm, N. Seiberg and E. Witten, \PLB{156}{85}{55}.}
\REF\WNP{S. Ferrara, N. Magnoli, T. Taylor and G. Veneziano, \PLB{245}{90}
{409}.}
\REF\LT{D. L\"ust and T. Taylor, \PLB{253}{91}{335}.}
\REF\EA{A. E. Faraggi and E. Halyo, preprint IASSNS-HEP-94/17, hep-ph/9405223.}
\REF\MOD{\AEF, \PLB{278}{92}{131}; \NPB{387}{92}{239}, hep-th/9208024.}
\REF\FFF{I. Antoniadis, C. Bachas, and C. Kounnas, \NPB{289}{87}{87};
I. Antoniadis and C. Bachas, \NPB{298}{88}{586};
H. Kawai, D.C. Lewellen, and S.H.-H. Tye, Nucl. Phys. B {\bf 288} (1987) 1.}
\REF\SQCD{D. Amati {\it et. al.}, \PRT{162}{88}{169}.}
\REF\KP{I. Antoniadis, J. Ellis, E. Floratos, \DVN ~and T. Tomaras,
\PLB{191}{87}{96}; S. Ferrara, L. Girardello, C. Kounnas, M. Porrati,
\PLB{192}{87}{368}; \PLB{194}{87}{358}.}
\REF\NANO{J. L. Lopez, D. V. Nanopoulos and K. Yuan, preprint CTP-TAMU-14/94,
hep-th/9405120.}
\REF\DSW{M. Dine, N. Seiberg and E. Witten, \NPB{289}{87}{589};
J. J. Atick, L. J. Dixon and A. Sen, \NPB{292}{87}{109};
	 S. Cecotti, S. Ferrara and M. Villasante, \IJMP{2}{87}{1839}.}
\REF\KLN{S. Kalara, J. L. Lopez and D. V. Nanopoulos,
\PLB{245}{91}{421}; \NPB{353}{91}{650}.}
%\REF\Louis{For a review see, J. Louis, SLAC-PUB-5645, DPF Conference, 1991.}

\singlespace
\rl{WIS--94/29/JUL--PH}
\rl{\today}
\rl{T}
\pagenumber=0
\normalspace
\smallskip
\titlestyle{\bf{Supersymmetry Breaking by Hidden Matter Condensation in
Superstrings}}
\smallskip
\author{Edi Halyo{\footnote\dag{e--mail address: jphalyo@weizmann.bitnet}}}
\smallskip
\centerline {Department of Particle Physics,}
\centerline {Weizmann Institute of Science}
\centerline {Rehovot 76100, Israel}
\vskip 2 cm
\titlestyle{\bf ABSTRACT}

We show that supersymmetry can be broken mainly by hidden matter condensates
in the observable matter direction in generic superstring models. This happens
only when the fields whose VEVs give masses to hidden matter do not decouple at
the condensation scale. We find how the parameters of the
string model and the vacuum determine whether supersymmetry is broken mainly by
hidden matter or gaugino condensates and in the matter or moduli directions.

\singlespace
\vskip 0.5cm
\endpage
\normalspace

\centerline{\bf 1. Introduction}

One of the most important but least understood aspects of superstring theories
is supersymmetry (SUSY) breaking. It is well--known that SUSY must be broken
non--perturbatively and around the $TeV$ scale in the observable sector
(to solve the hierarchy problem). Due to our lack of understanding of
non--perturbative string effects, the best we can do is to investigate
SUSY breaking
by non--perturbative phenomena in field theories which are low--energy
(i.e. $E<<M_P$) limits of superstring theories (such as string induced
supergravity [\SUGRA]).

The most common way of achieving dynamical SUSY breaking in superstrings
is by hidden gaugino condensation in supergravity [\GCM] theories which are
obtained from the massless sector of superstrings. In this
scenario, hidden sector gaugino condensates form when a non--Abelian hidden
gauge group becomes strong at a hierarchically small scale, $\Lambda_H<<M_P$.
The presence of such hidden sectors with non--Abelian gauge groups is a generic
feature of superstring models. The condensation is taken into account by a
non--perturbative superpotential, $W_{np}$, [\WNP] which has all the required
symmetry properties. One then finds that SUSY can be broken in the moduli
direction and in a
phenomenologically acceptable way (with the well--known problems of the
vanishing cosmological constant and the stability of the dilaton potential).

Most hidden sectors of superstring models also contain hidden matter
in the vector representations of the hidden gauge groups which condense with
the gauginos. Their presence not only affects the running of the hidden
gauge coupling constant but also modifies $W_{np}$
[\WNP,\LT]. In addition, matter condensation
can also be the source of SUSY breaking. Surprisingly, SUSY breaking by
hidden matter condensation has not attracted much attention until recently
[\EA]. In Ref. (\EA), the effect of hidden matter condensation on
$F$ terms was
examined in the framework of a realistic string model by examining the hidden
matter mass terms. In the following we will often
refer to the model of Ref. (\EA) as a concrete example of our results.
In this letter, we generalize the results of Ref. (\EA) by including the
effects of gaugino condensation and the full-fledged $W_{np}$ for superstrings.
We argue that SUSY can be broken by hidden matter rather than hidden
gaugino condensates and in the observable matter direction rather than the
moduli directions. We show that
this is a realistic possibility under quite generic conditions if the fields
whose VEVs give masses to hidden matter do not decouple at the condensation
scale. Our aim here is only to show the possibility of
this new kind of SUSY breaking in generic superstring models.
Whether this is the case or not in a specific
string model depends on the details of the model such as the hidden sector
gauge group and matter, the hidden matter mass terms etc. as we will show
below. This can only
be investigated in the framework of a specific model and with a detailed
numerical analysis of the scalar potential which we defer to the future.

\bigskip
\centerline{\bf 2. SUSY breaking by hidden gaugino condensation}

In this section we briefly review the gaugino condensation scenario
in superstrings. We consider a superstring model with a number of
generic properties to be outlined below rather than a specific one (such as
standard--like superstring models [\MOD]) for two reasons. First this makes
the discussion about SUSY breaking in realistic superstring models more
general. Second the properties we outline below (and in the next section)
can be seen loosely as necessary conditions for supersymmetry breaking by
hidden matter condensation .
We consider a superstring model in the four dimensional free fermionic
formulation [\FFF] with the following properties:

(a) The massless spectrum of the superstring model is divided into observable
and hidden sectors.
The observable sector contains a large number of states ($\phi_i$)
which are Standard Model (SM) singlets coming from the Neveu--Shwarz and some
twisted sectors in addition to the chiral generations.

(b) The hidden sector contains one (or more) $SU(N)$ non--Abelian gauge
group(s) with $M$ copies of matter ($h_i, \bar h_i$) in the vector
representations $N+\bar N$. The hidden matter states obtain masses
from non--renormalizable terms, $W_n$, in
the superpotential. Thus, the hidden matter mass matrix is non--singular
and the SUSY vacuum is stable [\SQCD]. $M<3N$ so that the hidden gauge group
is asymptotically free and condenses at the scale $\Lambda_H \sim M_v
exp(8 \pi^2/bg^2)$ where $b=M-3N$.

(c) The Kahler potential is generically given by [\KP]
$$K(S, S^\dagger, T, T^\dagger, \phi_i,  \phi_i^\dagger)=-log(S+S^\dagger)-
3log(T+ T^\dagger)-\sum_i \phi_i \phi_i^\dagger, \eqno(1)$$
where $S,T$ and $\phi_i$ are the dilaton, (overall) modulus and matter fields
respectively. These fields are in the ``supergravity basis" and are related to
the massless string states by well--known transformations. (For a recent
discussion of moduli and Kahler potentials in free fermionic models see Ref.
(\NANO).)

(d) The string vacuum is supersymmetric at the Planck scale, $M_P$. This is
guaranteed by satisfying the F and D constraints obtained from the cubic
superpotential $W_3$ (which is trilinear in $\phi_i$ and $h_i$) and the local
charges of the states. As we will see below, $W_3$ does not get any higher
order corrections as long as the hidden gauge group does not condense at
$\Lambda_H<<M_P$. Therefore, $W_3$ is the exact superpotential until hidden
sector condensation which results in SUSY breaking. The set of F and D
constraints is given by the following equations [\DSW]:
$$\eqalignno{&D_A=\sum_i Q^A_i \vert \l \phi_i \r\vert^2={-g^2e^{\phi_D}
\over 192\pi^2}Tr(Q_A) {1\over {2\alpha^{\prime}}}, &(2a) \cr
%&D^{\prime j}=\sum_k Q^{\prime j}_k \vert \chi_k \vert^2=0 \qquad j=1
%\ldots 5  {\hskip .1cm},&(11b) \cr
&D^j=\sum_i Q^j_i \vert \l \phi_i \r\vert^2=0 ,&(2b) \cr
&\l W_3 \r=\l{\partial W_3\over \partial \phi_i} \r=0 , &(2c) \cr}$$
where $\phi_i$ are the matter fields and $Q^j_i$ are their local charges.
$\alpha^{\prime}$ is the string tension
given by $(2\alpha^{\prime})^{-1}=g^2M_P^2/32\pi=g^2M_v^2$
and $Tr(Q_A)\sim 100$ generically in realistic string models. Eq. (2a) is the
D constraint for the anomalous $U(1)_A$ which is another generic
feature of realistic string models [\DSW]. We see that some
SM singlet scalars must get Planck scale VEVs of $O(M_v/10)$ in order to
satisfy Eq. (2a)
and preserve SUSY around the Planck scale. Then, due to the other F and D
constraints most of the other SM singlet scalars also obtain VEVs
of $O(M_v/10)$.

(e) There are non--renormalizable (order $n>3$) terms in the superpotential
which are generically of the form
$$W_n=c_n g^{n-2} h_i \bar h_j \phi_{j_1} \phi_{j_2} \ldots \phi_{j_{n-2}}
\eta(T)^{2n-6} M_v^{3-n}, \eqno(3)$$
which are obtained from the world--sheet correlators $A_n \sim \l V_1^f V_2^f
V_3^b \ldots V_n^b \r$ using the rules of Ref. (\KLN). $c_n$ are numerical
coefficients of $O(1)$ and $\eta$ is the Dedekind eta function. The powers of
$\eta$ and $M_v$ are such that the term $W_n$ has modular weight $-3$ and
dimension $3$ as it should. ($\phi_i$ and $h_i$ have modular weight $-1$
which is a generic feature of realistic models in the free fermionic
formulation.)
These terms contain both observable and hidden sector states. Once
the fields $\phi_i$ get VEVs, they give masses to the hidden states $h_i,
\bar h_i$. Therefore all the $n>3$ terms in Eq. (3) can be seen as hidden
matter mass terms. These corrections to $W_3$ when they become non--zero, (i.e.
when hidden matter condensates $\Pi_{ij}=h_i \bar h_j$ form) give corrections
to the cubic level F constraints in Eq. (2a) and destabilize the original SUSY
vacuum as was shown in Ref. (\EA).
(In general, there can also be terms of the form $c_n \phi_{i_1} \phi_{i_2}
\ldots \phi_{i_n}$, i.e. non--renormalizable terms with only observable fields.
These vanish in standard--like models [\EA] and we assume that they are not
present in the following.)

The above assumptions about the string model are relatively mild since
they are all generic features of realistic superstring models such as
standard--like models [\MOD].
We investigate what happens when the hidden gauge group condenses at
$\Lambda_H$ in two different cases: (a) when $\phi_i$ are heavy, i.e.
$m_{\phi_i}>>\Lambda_H$ and they decouple at $\Lambda_H$, and (b) when
$\phi_i$ are light i.e. $m_{\phi_i}<\Lambda_H$ and they remain in the spectrum.
In both cases we assume that the hidden matter states $h_i, \bar h_i$
do not decouple from the
spectrum at $\Lambda_H$ (otherwise obviously there can only be gaugino
condensation).

Case (a) is the case previously investigated in superstring models [\LT].
When the hidden gauge group condenses at $\Lambda_H$, gaugino condensates
$Y^3$ and matter condensates $\Pi_{ij}=h_i \bar h_j$ form. The
non--perturbative effective superpotential
obtained from the Ward identities and modular invariance is
$$W_{np}={1 \over {32\pi^2}}Y^3 log\{exp(32\pi^2S)[c \eta(T)]^{6N-2M} Y^{3N-3M}
det \Pi \}-tr A \Pi, \eqno (4)$$
where c is a constant and $A$ is the hidden matter mass matrix given by
the $n>3$ terms in Eq. (3). The last term corresponds
to the sum of all the $n>3$ terms in Eq. (3). The observable matter fields
$\phi_i$
appear only in the mass matrix $A$. In the flat limit $M_P \to \infty$,
gravity decouples and one gets a globally SUSY vacuum at which (in addition to
Eqs. (2a-c))
$${\partial W_{tot} \over {\partial Y}}={\partial W_{tot} \over
{\partial \Pi}}=0, \eqno(5)$$
where $W_{tot}=W_3+W_{np}$. We can replace $W_{tot}$ in Eq. (5) by $W_{np}$
since $W_3$ does not contain $Y^3$ or $\Pi$. The $n>3$ terms, $W_n$ which are
the hidden matter mass terms, are already included in $W_{np}$ through
$tr A \Pi$. The solutions to Eq. (5)
are used to obtain the composite fields $Y^3$ and $\Pi$ in terms of
$S,T,A$
$${1 \over {32\pi^2}}Y^3=(32\pi^2e)^{M/N-1}[c \eta(T)]^{2M/N-6}[det A]^{1/N}
exp(-32\pi^2S/N), \eqno (6)$$
and
$$\Pi_{ij}={1 \over {32\pi^2}}Y^3 A^{-1}_{ij}. \eqno(7) $$
Eqs. (6) and (7) are used to eliminate the composite fields in $W_{np}$ and
then
$$W_{np}(S,T)=\Omega(S) h(T) [det A]^{1/N}, \eqno(8)$$
where
$$\eqalignno{&\Omega(S)=-N exp(-32\pi^2S/N), &(9a) \cr
             &h(T)=(32\pi^2e)^{M/N-1}[c \eta(T)]^{2M/N-6}. &(9b)}$$
%or in other words
%$$W_{np}=-{{N Y^3} \over {32 \pi^2}} \eqno(10)$$
In $W_{np}$ all the information about the matter condensates, $\Pi$, and the
observable fields $\phi_i$ is contained in the term $det A$.
When $m_{\phi_i}>> \Lambda_H$ and $\phi_i$ decouple, one simply substitutes the
VEVs $\l \phi_i \r$ obtained from the solution to the F and D constraints in
$det A$. $\phi_i$ are longer dynamical fields since at the scale $\Lambda_H$
these heavy fields cannot be excited but simply sit at their VEVs. In this
sense, $\phi_i$ are similar to
the composite fields $Y^3$ and $\Pi$ which are also eliminated from $W_{np}$.
All $\phi_i$ do is to give masses to the hidden matter states $h_i, \bar h_i$
through their VEVs. As a result, in this case the only effect of matter
condensates $\Pi_{ij}$ is to change the scale of the gaugino condensate $Y^3$
through $det A$.

It is well--known that $W_{np}$ above breaks SUSY in the modulus direction
(but not in the dilaton direction), i.e. $\l F_T \r \not=0$ (but $\l F_S \r
=0$) where [\SUGRA]
$$F_k=e^{K/2}(W_k+K_kW). \eqno(10)$$
The subscript denotes differentiation with respect to fields and $k=S,T$. Here
$W=W_{tot}$. The vacuum is obtained by minimizing the scalar potential [\SUGRA]
$$V=\sum_k |F_k|^2 (G_k^{~k})^{-1}-3e^K|W|^2, \eqno(11)$$
%$$V=|F_S|^2 G_{SS^\dagger}^{-1}+|F_T|^2 G_{TT^\dagger}^{-1}-3e^K|W|^2
%\eqno(12)$$
where $k=S,T$ and $G=K+log|W|^2$. We do not give explicit expressions for
$F_k$ since they are special cases of the ones we obtain in the next section
where we include the effects of matter condensation and observable fields
$\phi_i$ with $m_{\phi_i}<\Lambda_H$.

\bigskip
\centerline{\bf 3. SUSY breaking by hidden matter condensation}

In this section we consider case (b) mentioned above in which $m_{\phi_i}<
\Lambda_H$ and $\phi_i$ remain in the spectrum. Then, $\phi_i$ should be
treated as dynamical fields similar to $S$ and $T$ since they can be excited
due to their small masses. Now $W=W(S,T,\phi_i)$
where from Eq. (8) all the $\phi_i$ dependence is in the term $det A$ which
arises due to the matter condensates $\Pi_{ij}$. As a result, in addition to
$F_{S,T}$ one should also check whether $F_{\phi_i}$ vanishes or not in the
vacuum. Also, it may now be possible to break SUSY mainly by
hidden matter condensation rather than hidden gaugino condensation.

$det A$ is a product of mass terms given generically by Eq. (3). Thus without
any loss of generality, we can
assume that it has the form
$$det A=k S^{-r} \phi_i^{s_i} \eta(T)^t \qquad r,s,t>0, \eqno(12)$$
where the $S$ dependence is obtained from the relation $g^2=1/S$ (at the string
tree level and for level one Kac--Moody algebras).
$\phi_i$ denotes any matter field which appears in $det A$ and $s_i$ is its
power. k is a constant of $O(1)$ which is given by the product of the relevant
$c_n$ in Eq. (3). In fact, this is the form of $det A$ which was obtained
from the explicit model of Ref. (\EA)
with $r=7$, $t=22$ and $s_i=1,5$ depending on the field $\phi_i$.
(In general, $det A$ is a sum of terms like that in Eq. (12).)
We see that there is a new $S$ and $T$ dependence in $W_{np}$ due to $det A$.
Taking this into account, we find for the $F$ term in the dilaton direction
$$\eqalignno{F_S={e^{-\phi_i \phi_i^\dagger/2} \over {(S+S^\dagger)^{1/2}
(T+T^\dagger)^{3/2}}} &h(T) [det A]^{1/N} \cr
& \times \{\Omega_S-{\Omega \over (S+S^\dagger)}+\Omega (log [det A]
^{1/N})_S \}. &(13)}$$
The first two terms in the curly brackets are the usual ones coming from
gaugino condensation.
The last term gives the contribution of the matter condensates (through $det
A$)
to $F_S$. Assuming the above form for $det A$ we get
$${\partial (log [det A]^{1/N}) \over \partial S}={-r \over {NS}}.
\eqno(14)$$
Using Eq. (9a) for $\Omega(S)$ and the fact that $S \sim 1/2$ in order
to have gauge coupling unification around $10^{18}~GeV$, we find that the first
term in the curly brackets always dominates the other two for realistic values
of $r$ and $N$. For example, in the explicit example of Ref. (\EA), $N=5$ and
$r=7$ and therefore the gaugino part is larger than the matter part by a factor
of $\sim 100$. In other words, the
effect of matter condensates on $F_S$ is negligible.

For the $F$ term in the modulus direction we find
$$\eqalignno{F_T={e^{-\phi_i \phi_i^\dagger/2} \over {(S+S^\dagger)^{1/2}
(T+T^\dagger)^{3/2}}}& \Omega(S) [det A]^{1/N} \cr
& \times \{h_T-{3h \over (T+T^\dagger)}+h (log[det A]^{1/N})_T \}. &(15)}$$
As for $F_S$, the first two terms in the curly brackets  arise from gaugino
condensation whereas the last one comes from matter condensation. Here
$${\partial h(T) \over \partial T}=-{h(T) \over {4 \pi}}G_2(T), \eqno(16)$$
where $G_2$ is the second Eisenstein function given by
$$G_2(T)={\pi^2 \over 3}-8 \pi^2\sum_n \sigma_1(n)e^{-2\pi n T}, \eqno(17)$$
and we used
$${\partial \eta(T) \over \partial T}=-{\eta(T) \over {4\pi}}G_2(T).
\eqno(18)$$
On the other hand, the contribution of the matter condensates are given by
$${\partial (log[det A]^{1/N}) \over \partial T}=-{t \over {4\pi N}} G_2(T).
\eqno(19)$$
Contrary to the $F_S$ case, this may or may not be larger than
the gaugino condensate part depending on the VEV of the modulus,
$\l T \r$ and the parameters $t,N$ as we will see below in more detail.

Finally, the hidden matter condensates, through the term $det A$, induce an $F$
term in the observable matter direction, $\phi_i$
$$\eqalignno{F_{\phi_i}={e^{-{\phi_i \phi_i^\dagger}/2} \over {(S+S^\dagger)^
{1/2} (T+T^\dagger)^{3/2}}} &[\Omega(S) h(T) [det A]^{1/N} \cr
& \times \left({s_i \over {N \phi_i}}+{\phi_i^\dagger \over
M_v^2}\right)+(W_{3\phi_i}+K_{\phi_i}W_3)]. &(20)}$$
This is exactly the result obtained in Ref. (\EA) in which the effect of matter
condensation on $F_{\phi_i}$ due to hidden matter mass terms was examined.
The last two terms simply give the contribution coming from the cubic
superpotential which vanishes for the solution to the F and D constraints
before the hidden gauge group condensed.
%As in Ref. (\EA), in the following we assume that
%the cubic F and D flat solution to Eqs. (2a-c) is still the vacuum in
%the presence of $W_{np}$ and therefore these terms still vanish.
Since generically the F and D flat solutions give $\l \phi_i \r \sim M_v/10$
we see that for realistic values of $s$ and $N$ the first terms in both
paranthesis in Eq. (20) (which correspond to the $W_k$ pieces in $F_k$)
dominate the second
ones. $F_{\phi_i}$ obviously arises solely from matter condensation since its
origin is the hidden matter mass term $tr A \Pi$ in Eq. (4).

The $F$ terms obtained above should be evaluated in the vacuum i.e. at the
minimum of the scalar superpotential which is given by Eq. (11) (but now
with $k=S,T,\phi_i$)
%$$V=|F_S|^2 G_{SS^\dagger}^{-1}+|F_T|^2 G_{TT^\dagger}^{-1}+|F_{\phi_i}|^2
%G_{\phi_i \phi_i^\dagger}^{-1}-3e^K|W|^2 \eqno(22)$$
in order to find if SUSY is broken or not and in what direction in field space.
This requires a complete numerical investigation of the scalar potential, $V$
which we defer to the future since our aim is only to raise the possibility
of SUSY breaking by hidden matter condensates and in the observable matter
direction. Instead, we will try to answer the following general questions in
the
following.

(a) Can $\l F_S \r$, $\l F_T \r$ or $\l F_{\phi_i} \r$ be non--zero for
realistic values of $\l S \r$, $\l T \r$ and $\l \phi_i \r$?

(b) Can the matter condensate contribution dominate that of the gaugino
condensate in $\l F_S \r$ and/or $\l F_T \r$? (We remind that
$\l F_{\phi_i} \r$ arises solely from matter condensates.)

(c) For what range of VEVs (and parameters $N,s,t,r$ etc.) does one of the
$F$ terms dominate the others, e.g.
$\l F_T \r>>\l F_{\phi_i} \r$ or vice versa?

{}From the explicit form of $V$ in Eq. (11) it is easy to show that
$${\partial V \over \partial S} \propto \Omega_S-{\Omega \over {S+S^\dagger}}
-\Omega(log[det A]^{1/N})_S, \eqno(21)$$
which means that $\l F_S \r=0$ by using Eq. (13). This is the analog of the
well--known result in
the pure gaugino condensate case (without the last term due to matter
condensates).
On the other hand, as in the pure gaugino condensate case,
we find that $\l F_T \r \not=0$ in general. With respect to $\l F_{\phi_i} \r$,
it was
shown in Ref. (\EA) that this is always non--zero once $W_n$ or hidden matter
mass terms are taken into account. The reason is that, the $n>3$ terms give
corrections to $W_3$ which turn the modified F constraints into an
inconsistent set of equations. Thus, the new
set of F constraints cannot be solved simultaneously for any set of SM singlet
scalar VEVs. Therefore, under our general assumptions, we find that
$\l F_{\phi_i} \r \not=0$ always, i.e. SUSY is always broken (by some amount
which depends on the parameters of the model)
in the observable matter direction in addition to the moduli direction.

We have seen that $F_S$ and $F_T$ have contributions from gaugino and matter
condensates. What are the relative magnitudes of these two contributions?
For $F_S$ this is not a relevant question since $\l F_S \r=0$ as we saw above.
For $F_T$ the situation is complicated since the
result depends strongly on $\l T \r$ due to the term $G_2(T)$ in Eq. (15).
For large $\l T \r \sim 1$ (in units of $M_v$), $G_2(T) \sim 3$ and then the
gaugino condensate part is dominant through the second term in the curly
brackets in Eq. (15). Depending on $t$ and $N$, the matter condensate part
given by the third term in Eq. (15) may also be important. For the second and
third terms to be comparable, one needs $t \sim 12 N$ which is rather large.
For example in the realistic model examined in Ref. (\EA) with $t=22$ and
$N=5$,
and gaugino condensate dominates for large $\l T \r$.
But $G_2(T)$ is a very rapidly decreasing (increasing in
absolute value which is relevant for us) function of $T$. A numerical
analysis of $G_2(T)$ shows that already for $\l T \r \sim 0.1$, $G_2$ is large
enough (in absolute value) so that the matter condensate part may become larger
than the gaugino condensate part depending on $t$ and $N$.
In the example of Ref. (\EA) with $t=22$ and $N=5$, the matter condensate part
is
in fact larger than the sum of the gaugino condensate contributions for
$\l T \r <0.1$.
In the intermediate range $0.1 < \l T \r <1$ both contributions to $\l F_T \r$
are of comparable magnitude.

Finally, we would like to know when both $\l F_T \r$ and $\l F_{\phi_i} \r$
are non--zero which one dominates? This will give the direction of SUSY
breaking in field space. From Eqs. (13) and (15) we find the ratio
$${\l F_T \r \over \l F_{\phi_i} \r} \sim {3N \over s_i}{\l \phi_i \r \over
\l T+T^\dagger\r}, \eqno(22)$$
for large $\l T \r \sim 1$ and
$${\l F_T \r \over \l F_{\phi_i} \r} \sim {t \over 4 \pi s_i} {\l \phi_i \r
G_2(\l T \r) }, \eqno(23)$$
for small $\l T \r < 0.1$. We find that (for $\l \phi_i \r \sim M_v/10$) in
the first case $\l F_{\phi_i} \r> \l F_T \r$ for $3 s_i>N$ and vice versa.
For example, for $N=5$ if $s_i=1$ then $\l F_{\phi_i} \r< \l F_T \r$
whereas if $s_i=5$ then $\l F_{\phi_i} \r> \l F_T \r$.
In the second case $\l F_{\phi_i} \r > \l F_T \r$ for $s_i>2 t$ and vice versa.
Note that in this case both $\l F_T \r$ and $\l F_{\phi_i} \r$ arise mainly due
to matter condensates. In the example of Ref. (\EA), $t=22$ and $s_i=1,5$ and
therefore $\l F_{\phi_i} \r<< \l F_T \r$.

The SUSY breaking scale in the observable sector which is given by the soft
SUSY breaking masses or the gaugino
mass $m_{3/2}$, must be phenomenologically acceptable, i.e. $ \sim O(TeV)$.
Using
$$m_{3/2}=e^{\l K/2 \r} {\l W_{tot} \r \over M_v^2}, \eqno(24)$$
and Eqs. (1), (6) and (8) for $K$, $Y^3$ and $W_{np}$ respectively (and $W_3$),
this phenomenological  constraint can be translated to constraints on the
parameters $N,M,r,s_i,t$ etc. These parameters of the string
model not only should result in SUSY breaking either by hidden gaugino or
matter
condensation but also a SUSY breaking scale of $O(TeV)$ in the observable
sector. The example
considered in Ref. (\EA) with $N=5$, $M=3$, $r=7$, $t=22$ and $s_i=1,5$ gives
$m_{3/2} \sim 1~TeV$ as was shown by an explicit calculation.

\bigskip
\centerline{\bf 4. Conclusions and discussion}

In this letter, we have shown that under quite general assumptions SUSY
breaking
by hidden matter condensation in the observable matter direction is possible.
This should be compared with the conventional mechanism of breaking SUSY by
hidden gaugino condensates and in the moduli direction. We have shown that both
mechanisms are possible for a given string model. Whether one or the other
occurs depends on the details of the string model such as the hidden gauge
group, hidden matter content and the hidden matter mass terms and can only be
decided by a detailed analysis of a given model.

In addition to the quite general assumptions we made in Section 2, a necessary
condition for SUSY breaking by hidden matter condensation,
with $\l F_{\phi_i} \r\not=0$ is the following:
the observable fields $\phi_i$ whose VEVs give masses to the hidden matter
must not be heavier
than the hidden gauge group condensation scale, $\Lambda_H$. Otherwise,
$\phi_i$ decouple at $\Lambda_H$ and SUSY can only be broken by
hidden gaugino condensation. This condition puts severe constraints on the
F and D flat solution at $M_P$ since some VEVs $\l \phi_j \r$ must vanish so
as not to give masses of $O(M_v/10)$ to $\phi_i$ from $W_3$. We find that for
large $\l T \r \sim 1$, $\l F_{\phi_i} \r$ (due to matter condensates) can
be either larger or smaller than $\l F_T \r$ (due to gaugino condensates)
depending on $N$
and $s_i$. For small $\l T \r <0.1$ the hidden matter condensation mechanism is
dominant and $\l F_{\phi_i} \r <<\l F_T \r$. In the intermediate range
$0.1<\l T \r <1$, $\l F_{\phi_i} \r \sim \l F_T \r$ and matter and gaugino
condensates contributions to $\l F_T \r$ are comparable. $\F_{\phi_i}$ arises
solely from hidden matter condensation. We also find that $\l F_S \r=0$ as
in the pure gaugino condensate case.

Obviously, $\l T \r$ and the other VEVs such as $\l S \r$ and $\l \phi_i \r$
(i.e. the vacuum) are not arbitrary but are fixed dynamically by the
non--perturbative superpotential $W_{np}$. One should minimize the scalar
potential, V, given by Eq. (11) to find these VEVs in a given model. Since our
aim in this work was just to show the possibility of a new SUSY breaking
scenario, we did not investigate the scalar potential in detail. We have not
touched upon the of the dilaton stability and cosmological
constant problems which are closely connected to SUSY breaking either since
this too requires a dynamical determination of the VEVs.
In the future this should be done in the framework of realistic string models
such as the one investigated in Ref. (\EA).
The vacuum which is fixed dynamically together with the parameters of the
string
model will determine which SUSY breaking mechanism actually occurs in a given
model.

\bigskip
\centerline{\bf Acknowledgements}
This work was supported by the Department of Particle Physics and a Feinberg
Fellowship.

\vfill
\eject

\refout
\vfill
\eject

\end
\bye